\documentstyle[psfig]{aipproc}

\begin{document}
\title{Dielectric screening in doped Fullerides}
\author{Erik Koch$^{a,b}$, Olle Gunnarsson$^{a}$, and Richard M.~Martin$^{b}$} 
\address{
  $^{a)}$ Max-Planck-Institut f\"ur Festk\"orperforschung, 
         70569 Stuttgart, Germany\\
  $^{b)}$ Department of Physics and Materials Research Laboratory,\\ 
          University of Illinois, Urbana, IL 61801, USA}
\maketitle
\begin{abstract}
For conventional superconductors the electron-electron interaction is strongly
reduced by retardation effects, making the formation of Cooper pairs possible.
In the alkali-doped Fullerides, however, there are no strong retardation
effects.  But dielectric screening can reduce the electron-electron
interaction sufficiently, {\em if} we assume that the random-phase
approximation (RPA) is valid. It is not clear, however, if this assumption
holds, since the alkali-doped Fullerides are strongly correlated systems
close to a Mott transition. To test the validity of the RPA for these systems
we have calculated the screening of a test charge using quantum Monte Carlo.
\end{abstract}
  
\section*{Introduction}

In order to lead to an effective attraction between electrons, the 
electron-phonon interaction has to overcome the strong electron-electron
repulsion. For conventional superconductors retardation effects strongly
reduce the Coulomb repulsion \cite{sc}. The resulting effective interaction
is described by the dimensionless Coulomb pseudopotential $\mu^\ast$. 
In the above sense, the alkali-doped Fullerides are non-conventional 
superconductors, since for them retardation effects are very inefficient in 
reducing the Coulomb repulsion \cite{C60RPA}. One can, however, conceive an 
alternative mechanism for reducing $\mu^\ast$: Within the random phase
approximation (RPA) the Coulomb interaction is strongly reduced by the 
very efficient dielectric screening \cite{C60RPA}.

It is not clear, however, how well the random phase approximation describes
the screening in the alkali-doped Fullerides. RPA works well in the limit of
weak interaction, while in the opposite limit it is qualitatively wrong.
Very little is known about the screening in the intermediate region, where
interaction and kinetic energy are comparable. This is the region where the 
superconducting Fullerides, being close to a Mott transition, lie.

To find out how well RPA works for the alkali-doped Fullerides, we have 
studied the static screening of a test charge in a Hubbard-like model using
quantum Monte Carlo methods. Our results indicate that RPA gives a surprisingly
accurate description of the static screening on the metallic side of a Mott
transition.

\section*{Quantum Monte Carlo Calculations}

To describe the conduction electrons in A$_3$C$_{60}$ we use the multi-band
Hubbard-like Hamiltonian
\begin{displaymath}
  H_0 = \sum_{\langle i,j \rangle\; m,m',\sigma}  
          t_{ij\,mm'} c^\dagger_{im\sigma} c_{jm'\sigma} 
      + U\sum_{i,\;m\sigma<m'\sigma'} n_{im\sigma} n_{im'\sigma'} .
\end{displaymath}
The indices $i$, $j$ label the sites of the fcc lattice, $\langle i,j \rangle$
denoting nearest neighbors. Each C$_{60}$-molecule randomly takes one of two
orientations \cite{disorder}. The hopping matrix elements $t_{ij\,mm'}$ 
between the three-fold degenerate $t_{1u}$ orbitals on neighboring sites are 
obtained from a tight-binding parameterization \cite{TBparam}. The Hubbard 
interaction $U$ is varied to study the effect of correlations. Experimental 
estimates give $U\approx 1.5\,eV$ \cite{HubbardU}.

To study the static screening, we introduce a test charge $q$ on a site $i_q$.
The corresponding term in the Hamiltonian is
\begin{displaymath}
H_1(q) = q\,U\sum_{m\sigma} n_{i_q m\sigma} . 
\end{displaymath}
We can then find the response of the system upon introduction of the test 
charge by comparing the electron density $n_q$ on site $i_q$ for the system 
with the test charge, to the electron density $n_0$ for the unperturbed system:
$\Delta n = n_0 - n_q$.

We determine the ground state properties of the Hamiltonians $H_0$ and $H_0+H_1$
(for finite fcc clusters) using diffusion Monte Carlo with the fixed-node
approximation \cite{tenHaaf}. In this method we construct a trial wave function
$|\Psi_T\rangle$ and allow it to diffuse towards the exact (within the fixed 
node approximation) solution $|\psi_0\rangle$. For the trial function, we make
a generalized Gutzwiller ansatz:
$\langle R|\Psi_T\rangle = g^{d(R)} g_q^{n_q(R)}\;\langle R|\Phi\rangle$.
$|\Phi\rangle$ is the Slater determinant calculated using the random phase
approximation. The Gutzwiller factors reflect the two interaction terms, 
where for a given configuration $R$ of the electrons 
$d(R)=\langle R|\sum_{i,\,m\sigma<m'\sigma'} 
n_{im\sigma} n_{im'\sigma'}|R\rangle$ and $n_q(R)=\langle R|\sum_{m\sigma} 
n_{i_qm\sigma}| R \rangle$. We optimize the free parameters $g$ and $g_q$ 
using a correlated sampling technique.

Diffusion Monte Carlo only provides us with the mixed estimator 
$\langle\Psi_T|{\cal O}|\Psi_0\rangle / \langle\Psi_T|\Psi_0\rangle$
of an observable ${\cal O}$. If, however, the trial function $|\Psi_T\rangle$ 
is close to the ground state, we can estimate the ground state expectation 
value using the extrapolated estimator
\begin{displaymath}
  {\langle\Psi_0|{\cal O}|\Psi_0\rangle \over \langle\Psi_0|\Psi_0\rangle}
  \approx
   2\;{\langle\Psi_T|{\cal O}|\Psi_0\rangle\over\langle\Psi_T|\Psi_0\rangle} 
   -  {\langle\Psi_T|{\cal O}|\Psi_T\rangle\over\langle\Psi_T|\Psi_T\rangle} .
\end{displaymath}
The last term, the expectation value of the observable for the trial function,
is easily calculated using variational Monte Carlo. To make sure that the
extrapolated estimator gives reliable results, we have compared our Monte
Carlo calculations with the results from the exact diagonalization for small
systems. For larger systems, where exact diagonalization is not possible, we
have checked that the trial function is close to the ground state wave 
function.

\section*{Results and Discussion}

\begin{figure}
\centerline{\psfig{figure=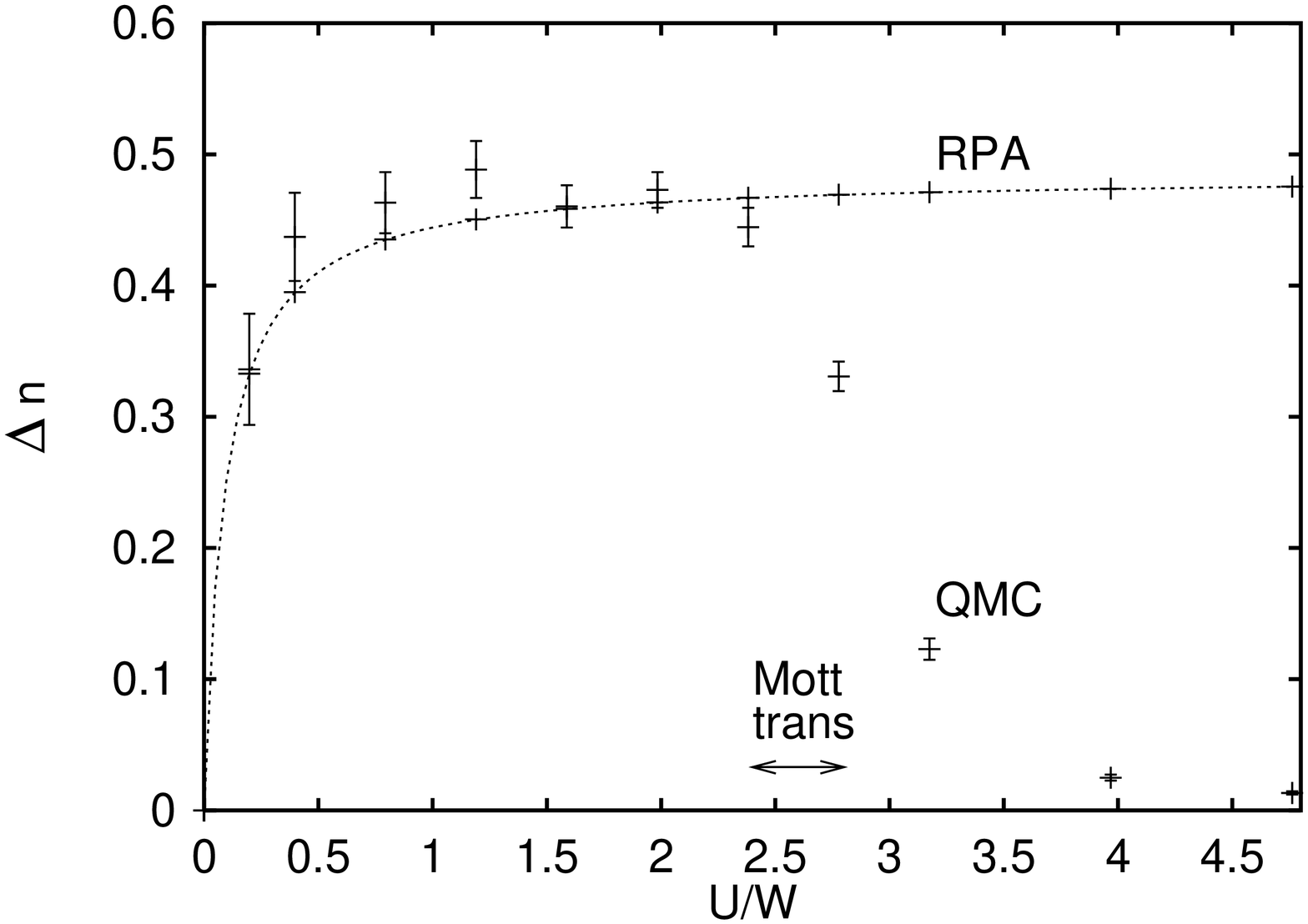,height=6.6cm}}
\caption{\label{screen}
 Screening of a test charge $q=0.5\,e$ as a function of U for a cluster
 of 32 molecules (band width $W=0.63\,eV$). The crosses (interpolated by the 
 dotted curve) show the results of a calculation using the random phase 
 approximation (RPA). The results of the quantum Monte Carlo (QMC) 
 calculations are given with their respective error bars. The region where 
 the system undergoes a Mott transition is indicated. The RPA screening 
 remains rather accurate almost all the way up to the Mott transition, but 
 fails badly in the insulating region.  
}
\end{figure}

We have calculated the screening charge $\Delta n$ for finite clusters
of different size. For a given test charge, all calculations give similar 
results. As an example, Fig.~\ref{screen} shows the screening of a test charge 
$q=0.5\,e$ as a function of the Hubbard interaction $U$ for a cluster of 32 
molecules. 

Looking at the results from the quantum Monte Carlo calculations, we can
distinguish two qualitatively different regions. For weakly interacting 
systems ($U/W$ small) the screening is very good, while it breaks down
completely for large $U/W$, where the system is a Mott insulator.
RPA somewhat underestimates the screening for $U/W \lesssim 1.0$ 
\cite{underest}. For intermediate values of $U/W$ it works surprisingly
well and actually remains rather accurate almost all the way up to the Mott
transition. Beyond this point RPA becomes qualitatively wrong.

The failure of the RPA in the strongly interaction regime can be easily
understood. In RPA it only costs kinetic energy to screen the test charge.
Since in the strongly interacting region $W/U$ is small, the cost in kinetic
energy is small compared to the interaction with the test charge. Therefore,
when a test charge is put on some site, RPA predicts that almost the same 
amount of electronic charge leaves that site in order to minimize the 
interaction with the test charge. Such a calculation neglects, however, the
fact that in a strongly correlated system the electrons that leave the site 
with the test charge have to pay a large correlation energy. In the QMC 
calculations this is taken into account properly, leading to the observed 
breakdown in the screening for large $U/W$.

The gradual decrease in the efficiency of the screening just before the Mott 
transition that can be seen in Fig.~\ref{screen} suggests a mechanism for 
understanding the anomalous behavior of NH$_3$K$_3$C$_{60}$. At normal pressure
this system is an insulator. Under pressure it becomes superconducting, with
$T_c$ increasing with pressure \cite{NH3K3C60}. This is in contrast to the 
other alkali-doped Fullerides, where $T_c$ drops with pressure. As pressure is
applied to the Mott insulator, $U/W$ decreases and the screening becomes more
and more efficient, reducing the effective Coulomb interaction $\mu^\ast$, i.e.
increasing $T_c$. As $U/W$ further decreases, the screening saturates. Now the
reduction of the electron-phonon interaction (because of the decreasing density
of states) dominates, and the standard behavior of the alkali-doped Fullerides 
is recovered. 

To summarize, our results indicate that RPA is surprisingly accurate almost
all the way up to the Mott transition. Since retardation effects are 
inefficient, this has important implications for the superconductivity in the 
alkali-doped Fullerides.

\section*{Acknowledgments}
This work has in part been supported by the Department of Energy (grant 
DEFG 02-91ER45539). E.K.\ thanks the Alexander-von-Humboldt-Stiftung for 
financial support under the Feodor-Lynen-Program.

\end{document}